\documentclass[prl,showpacs,superscriptaddress,twocolumn]{revtex4}
\usepackage{graphicx}
\usepackage{float}
\usepackage{amsmath}
\usepackage{amssymb}
\usepackage{bm}

\newcommand{\rmd}{{\rm d}}
\newcommand{\rme}{{\rm e}}
\newcommand{\rmi}{{\rm i}}
\newcommand{\sign}{\operatorname{sgn}}

\begin{document}

\title{Theory of the quantum Hall effect in graphene}
\author{Tobias~Kramer}
\affiliation{Institute for Theoretical Physics, University of Regensburg, 93040 Regensburg, Germany}
\affiliation{Department of Physics, Harvard University, Cambridge, MA~02138, USA}
\author{Christoph~Kreisbeck}
\author{Viktor Krueckl}
\affiliation{Institute for Theoretical Physics, University of Regensburg, 93040 Regensburg, Germany}
\author{Eric~J.~Heller}
\affiliation{Department of Physics, Harvard University, Cambridge, MA~02138, USA}
\affiliation{Department of Chemistry and Chemical Biology, Harvard University, Cambridge, MA~02138, USA}
\author{Robert~E.~Parrott}
\affiliation{Department of Physics, Harvard University, Cambridge, MA~02138, USA}
\affiliation{School of Engineering and Applied Science, Harvard University, Cambridge, MA~02138, USA}
\author{Chi-Te~Liang}
\affiliation{Department of Physics, National Taiwan University, Taipei, Taiwan 106, R.O.C.}

\begin{abstract}
We study the quantum Hall effect (QHE) in graphene based on the current injection model. In our model, the presence of disorder, the edge-state picture, extended states and localized states, which are believed to be indispensable ingredients in describing the QHE, do not play an important role. Instead the boundary conditions during the injection into the graphene sheet, which are enforced by the presence of the Ohmic contacts, determine the current-voltage characteristics.
\end{abstract}

\pacs{73.63.-b,73.21.-b,73.43.-f}

\maketitle

Several experiments have studied the quantum Hall effect (QHE) in graphene \cite{Novoselov2005a,Zhang2005a,Zhang2006a}. Recently the observation of the fractional QHE in a suspended high-mobility graphene device has been reported using a two-terminal measurement setup \cite{Du2009a,Bolotin2009a}. Surprisingly, four-terminal measurements in small graphene devices did not reveal the QHE. This observation has been related to the large influence of the metallic contacts on the formation of the Hall potential in small graphene flakes, where the current and voltage probes short out the Hall device \cite{Du2009a}. For the theory of the QHE, the experiments provoke fundamental questions about the role of the metallic contacts, the correct incorporation of boundary conditions, and the importance of electron-electron interactions in graphene. These questions are not addressed in the commonly used edge-state and disorder models of the QHE, which have been proposed for the QHE in graphene and have been taken over from models for conventional semiconductors \cite{Koshino2007a}. Here, we give further theoretical considerations which stress the importance of boundary effects in small Hall devices and the role of electron-electron interactions for the integer and fractional QHE.\\
Before we discuss the boundary effects, let us see why boundary effects are in general not considered to be part of theories of the QHE. Two theoretical models are commonly invoked to explain the IQHE: the disorder model and the edge-state picture \cite{Ando2003a}. The disorder model adapts a specific distribution of the density of states (DOS) of a quasi two-dimensional electron gas in the presence of a strong perpendicular magnetic field and a randomly fluctuating potential. It is assumed that the DOS splits into two parts centered around each Landau level: an extended state band at the center of each Landau level which is bordered by a broad region of localized states. These results are obtained by using ensemble averaged Green functions representing an infinite two-dimensional system \cite{Prange1987a}. The QHE is viewed as a phase transition, which is not related to the measurement apparatus with its contacts.
\begin{figure}[b]
\begin{center}
\includegraphics[width=0.4\textwidth]{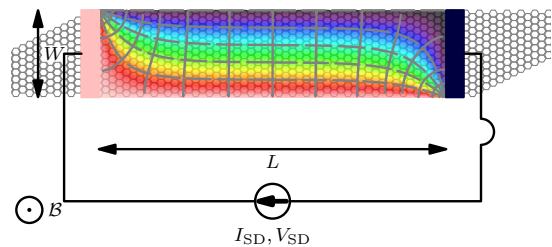}
\end{center}
\caption{Hall potential (color online) obtained by solving the Laplace equation under the Hall boundary conditions of a steady current flow in the presence of a strong magnetic field. At the upper left corner electrons enter the device in a region of high electric fields and move to the lower right corner.
\label{fig:HallBar}}
\end{figure}
The second model of the QHE, the edge-state approach, restricts the two-dimensional plane by two edges between semi-infinite leads.  The strong magnetic field results in a quasi one-dimensional transport along the two edges and thus gives rise to two oppositely flowing currents. A four terminal measurement should guarantee a clean signature of the QHE even in the presence of disorder within the device \cite{Buettiker1988a}.
Both models use contrived translational invariances which are not present in an actual Hall device. The metallic nature of the Ohmic contacts at the device border and the specific sample geometry play no special role and are ignored. This is not the case for the classical Hall effect, where the boundary conditions and the device geometry are crucial for the calculation of the self-consistent Hall potential \cite{Wick1954a,Thompson1969a,Moelter1998a}. For the determination of the classical Hall potential, the metallic contacts have to be considered as equipotential surfaces, which enforce also in the two-dimensional subsystem a uniform potential underneath the contacts \cite{Kramer2009c}. The two-terminal resistance of a classical Hall device can be readily calculated by the ratio of the source-drain voltage to the source-drain current. The specific device geometry and the placement of metallic contacts do have a strong influence on the Hall potential solution \cite{Kramer2009c}. The fundamental reason for the formation of the Hall potential are the interactions between the electrons in the complete device, including the contacts, which translate the magnetic Lorentz force acting on every electron into a global, non-trivial adjustment of the potential with the emergence of two hot-spots at opposite corners of the device. The interactions are not present in a Fermi liquid model of effectively non-interacting electrons \cite{Prange1987a} and thus these models are not sufficient to explain the experimental observations of hot-spots and the emergence of the classical Hall field in the QHE.\\
In the following we describe the injection model of the QHE in graphene for a two-terminal measurement. We incorporate the metallic contacts in the model and thus employ different boundary conditions than either the edge-state or disorder models. Our perspective is analogous to the theory of scanning tunneling microscopy and quantum point contacts, in which electron flow in a restricted region ultimately determines the conductance. The mean field potential (Fig.~\ref{fig:HallBar}) gives rise to an injection hotspot and an exit hotspot, where the drift velocity takes on its larges value due to the fast change in the Hall potential over a small corner region. Interestingly, images of the Hall potential in GaAs/AlGaAs heterostructures in Ref.~\cite{Knott1995a} closely resemble Fig.~\ref{fig:HallBar}. The rate of electrons entering the device within the hotspot region is given (as in the edge-state model) by the convolution integral of the LDOS at the injection points with the group velocity and the Fermi-Dirac distribution. The main difference to an edge-state model is the absence of translational invariance due to the presence of the contacts and the strong bias of the current injection towards a corner of the device. The resulting current flows unidirectional diagonally across the device.\\
From the classical solution alone, no reason for the quantization of the resistivity exists. The quantization requires to study the relation between the source-drain voltage and the current in a Hall bar. The self-consistent potential in a Hall device does depend on the applied voltages, the magnetic field, and the geometry of the sample. For rectangular samples with a length to width aspect ratio $L/W \gg 1$, the solution of the Laplace equation for a Hall bar in the presence of a magnetic field and current leads to a geometry independent solution with high electric field in two opposite corners of the device \cite{Thompson1969a}, where the Hall potential attains the universal form
\begin{equation}\label{eq:Vcorner}
V_{\text{corner}}(x,y)=\frac{2}{\pi}V_{\mathrm{SD}}\arctan(y/x).
\end{equation}
Here, $V_{\mathrm{SD}}$ denotes the voltage difference between the source and drain contacts. For the inverse geometrical ratio, $L/W\ll 1$ (used for measuring the FQHE in graphene \cite{Du2009a}), a very similar solution emerges, since the long contact region enforce the boundary conditions very efficiently.
In the following, we will use the uniform-map solution as the mean-field potential, which emerges due to interaction and screening between the electrons and the positive charges and by considering the metallic boundary conditions at the current source and sinks \cite{Kramer2009c}. We study the propagation of the effectively non-interacting electrons in the mean-field potential. We view our model as providing a starting point for a more rigorous inclusion of interaction effects, already in the non-fractional quantum Hall effect. Experimental evidence for the existence and relevance of the mean-field solution is provided by the absence of the QHE in a four-terminal measurement in graphene, which has been attributed to the dominance of the metallic boundary conditions at the contacts in small devices \cite{Du2009a}.
For low energies, the nearest-neighbor tight-binding Hamiltonian of graphene can be expressed by an effective Dirac-type Hamiltonian \cite{Koshino2007a}. Including the minimal coupling of the potential $V(x,y)$ and the magnetic field via the vector potential $\boldsymbol{{\cal A}}= {\cal B} (-y,0,0)$ to the kinematic momentum $\boldsymbol{\Pi}=\mathbf{p}-e\boldsymbol{{\cal A}}$ yields:
\begin{equation}
H =c
\left(\begin{array}{cccc}
V(x,y)/c & \Pi_x-\rmi \Pi_y & 0 & 0\\
\Pi_x+\rmi \Pi_y & V(x,y)/c & 0 & 0\\
0 & 0 & V(x,y)/c & \Pi_x+\rmi \Pi_y \\
0 & 0 &\Pi_x-\rmi \Pi_y & V(x,y)/c
\end{array}\right).
\end{equation}
The four components of the wavefunction are labelled $(\psi_1,\psi_2,\psi_3,\psi_4)=(\psi_A^K,\psi_B^K,\psi_A^{K'},\psi_B^{K'})$, where $K$ and $K'$ refer to the two $K$ points in the first Brillouin zone and the two sublattices $A$ and $B$, and $c\approx 10^6$~m/s denotes a velocity. 
The Hamiltonian omits the spin degree of freedom, which is added later using an effective $g$-factor.
\begin{figure}[b]
\begin{center}
\includegraphics[width=0.4\textwidth]{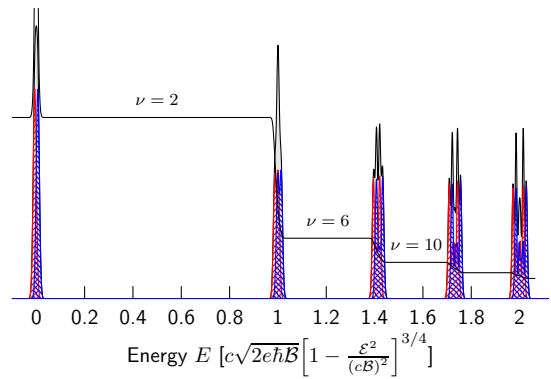}
\end{center}
\caption{
Overview of the LDOS (color online) with an effective $g$-factor $g^{*}=2$ for the electric field ${\cal E}=200$~kV/m and the magnetic field  ${\cal B}=15$~T as a function of the energy in Eq.~(\ref{eq:ldos}). The red and blue areas are the contributions from the spin-up and spin-down components. Note the modulation within each spin-split component, which is caused by the electric field and increases for higher levels. The black solid line shows $\rho_{xy}=\frac{h}{e^2\nu}$. The energy is given in units of $141$~meV.
\label{fig:ldos_overview}}
\end{figure}
The propagation of wave packets by numerical methods allows us to accurately determine the local density of states (LDOS) without the need to introduce half-infinite leads, which are not compatible with the boundary conditions discussed above. We have adapted the approach of Ref.~\cite{Kramer2008a} to the massless Dirac equation in graphene using a recursively evaluated polynomial expansion of the time evolution operator \cite{Krueckl2009a}. We calculate the LDOS in a strong magnetic field and for the potential given in Eq.~(\ref{eq:Vcorner}) by tracking the time-dependent autocorrelation function for several picoseconds. The resulting LDOS is intrinsically broadened (by decay in the autocorrelation function due to flux leaving the injection area under the combined influence of the electric field potential and the magnetic field) and shows only extended states which connect one corner of the device with the opposite one. We find that our numerical results for the LDOS in the corner potential (\ref{eq:Vcorner}) are in excellent agreement with the analytically derived LDOS in perpendicular and homogeneous electric and magnetic fields, provided we choose the homogeneous electric field value to match the local potential gradient:
\begin{equation}\label{eq:Ecorner}
{\cal E}(\mathbf{r})=\left|-\frac{\nabla V_{\text{corner}}(x,y)}{e}\right|=\frac{2}{\pi}\frac{V_{\mathrm{SD}}}{r}.
\end{equation}
The uniform field case is analytically solvable using the proper-time approach of Fock~\cite{Fock1937a,Lam1970a,Lam1971a,Dodonov1976a}.
We construct the LDOS in the magnetic field $\boldsymbol{{\cal B}}=(0,0,{\cal B})$ and the electric field $\boldsymbol{{\cal E}}=(0,{\cal E},0)$ from the four scalar components and the eigenenergies 
\begin{equation}
E_{n,p_x}
=c \sign(n)\sqrt{2|n|e\hbar {\cal B}} {\left[1-\frac{{\cal E}^2}{{(c \cal B)}^2}\right]}^{3/4}+\frac{{\cal E}}{{\cal B}} p_x,
\end{equation}
given in Landau gauge  \cite{Nieto1985a}. The spinor eigenfunctions can be written in terms of oscillator functions
\begin{equation}
u_n(\xi)=\frac{\rme^{-\xi^2/2} {\rm H}_{n}(\xi)}{\sqrt{2^n n! \sqrt{\pi}}},
\end{equation}
\begin{equation}
\xi=\sqrt{\frac{e\beta}{\hbar c}}\big(y+\frac{c^2 p_x {\cal B} - {\cal E} E_{n,p_x} }{e \beta^2}\big),\;\beta=\sqrt{{(c \cal B)}^2-{\cal E}^2}.
\end{equation}
Using the coefficients
\begin{equation}
a=-\frac{ \sqrt{c{\cal B}+\beta}}{\sqrt{2{c\cal B}}},\quad
b=\frac{{\cal E}}{\sqrt{2{c\cal B}}} \frac{1}{ \sqrt{c{\cal B}+\beta} },
\end{equation}
with the plane wave solution in the $x$-direction,
$\phi_{p_x}(x)=\rme^{\rmi p_x x/\hbar}/\sqrt{2\pi}$,
we obtain
\begin{equation}
\psi_{1,n,p_x}(\mathbf{r})=\phi_{p_x}(x){\left(\frac{e\beta}{\hbar c}\right)}^{1/4}(\sign(n)\;a\; u_{|n|}(\xi)-b\; u_{|n|-1}(\xi)),
\end{equation}
\begin{equation}
\psi_{2,n,p_x}(\mathbf{r})=\phi_{p_x}(x){\left(\frac{e\beta}{\hbar c}\right)}^{1/4}(a\;u_{|n|-1}(\xi)-\sign(n)\;b\; u_{|n|}(\xi)),
\end{equation}
\begin{equation}
\psi_{3,n,p_x}(\mathbf{r})=\psi_{2,n,p_x}(\mathbf{r}),\quad \psi_{4,n,p_x}(\mathbf{r})=\psi_{1,n,p_x}(\mathbf{r}).
\end{equation}
For $n=0$, the $u_{|n|-1}(\xi)$ terms vanish and the wavefunctions acquire an additional normalization factor of $\sqrt{2}$.
We obtain the LDOS by evaluating
\begin{eqnarray}
n_{{\cal E}\times{\cal B}}(\mathbf{r};E)
&=&\sum_{i=1}^4 \sum_{n=-\infty}^{\infty}\int \rmd p_x {|\psi_{i,n,p_x}(\mathbf{r})|}^2 \delta(E-E_{n,p_x})\notag\\
&=&\sum_{i=1}^4 \sum_{n=-\infty}^{\infty} {\left|\frac{\partial E_{n,p_x}}{\partial p_x}\right|}_{p_x=p_{\delta}}^{-1} {|\psi_{i,n,p_\delta}(\mathbf{r})|}^2,
\end{eqnarray}
\begin{equation}
p_{\delta}=\frac{{\cal B}}{{\cal E}}E-\sign(n)\frac{{c \cal B}}{{\cal E}}\sqrt{2|n|e\hbar\cal B}{\left[1-\frac{{\cal E}^2}{{(c \cal B)}^2}\right]}^{3/4}.
\end{equation}
In the presence of an electric field, the spinor index $i$ can no longer be used to identify the sublattice. The introduction of the real spin completes the computation of the LDOS
\begin{equation}\label{eq:ldos}
n^{\uparrow\downarrow,A,B}_{{\cal E}\times{\cal B}}(\mathbf{r};E) =
n_{{\cal E}\times{\cal B}}\big(\mathbf{r};E - \frac{g^*\mu_B}{2}\big)+
n_{{\cal E}\times{\cal B}}\big(\mathbf{r};E + \frac{g^*\mu_B}{2}\big),
\end{equation}
where $\mu_B$ denotes the Bohr magneton (which contains the normal electron mass, not the effective one) and $g^*$ is the effectice $g$-factor of the electron. The LDOS shown in Fig.~\ref{fig:ldos_overview} is symmetric with respect to the $E=0$ value. The centers of the spin-split LDOS are located at energies $E_{n,p=0} \pm \frac{1}{2}g^*\mu_B$.
For electric fields approaching ${\cal E}/{\cal B}=c$, the harmonic oscillator functions are replaced by Airy functions \cite{Lam1971a}. The transition from a magnetic field dominated LDOS to an electric field dominated one for the non-relativistic case is also discussed in Ref.~\cite{Kramer2003a}. The group velocity $\partial E_{n,p_x}/\partial p_x$ in the relativistic case is given by ${\cal E}/{\cal B}$ and thus unchanged in the non-relativistic limit and does not depend on the energy of the particle. 
\begin{figure}[b]
\begin{center}
\includegraphics[width=0.41\textwidth]{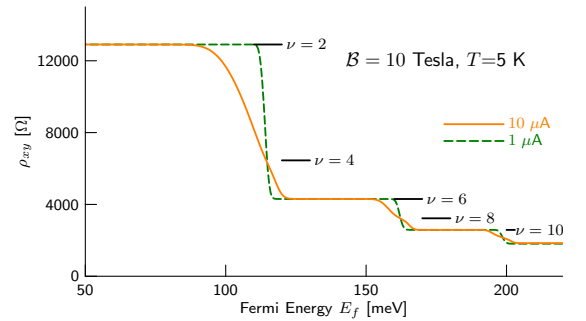}
\end{center}
\caption{
Hall resistivity as function of Fermi energy (color online). Shown are the filling factors $\nu=2-10$ at a temperature of $T=5$~K for two dc-currents $I_{SD}=1\;\mu$A and $I_{SD}=10\;\mu$A.
\label{fig:CC}}
\end{figure}
In the injection model, the LDOS in the injection region together with the group velocity determines the injected current density. The magnitude of the electric field in the middle of the device is given by $V_{\mathrm{SD}}/W$ and thus a higher source-drain voltages increases the drift velocity. An important quantity is the statistical distribution of the electric field strengths at the injection sites. Since fully quantum-mechanical simulations of open quantum-systems with Coulomb interactions are not available, we have to make assumptions about the magnitude and spatial distribution of the electric field values. Also dissipative processes are expected to occur at the hot-spots at the two opposite corners with the highest probability for electrons to enter and to leave the graphene flake. The superposition of various field values leads to a suppression of the electric field induced gaps within the Landau levels. The current is given by the product of the group velocity and the LDOS convoluted with the temperature-dependent Fermi-Dirac distribution $f$:
\begin{equation}\label{eq:IHall}
I_{SD}=e\int_0^{\infty}\rmd E \int \rmd\mathbf{r_c}\; f(E-E_F) 
\frac{{\cal E}(\mathbf{r}_c)}{{\cal B}}
n^{\uparrow\downarrow,A,B}_{{\cal E}\times{\cal B}}(\mathbf{r}_c;E),
\end{equation}
where the magnitude of the electric field in the injection region near the hot-spot depends on the value of the source-drain voltage drop occurring in this corner, given by Eq.~(\ref{eq:Ecorner}). Thus we obtain a voltage dependent group velocity and broadening of the LDOS, depending on the injection points at $\mathbf{r}_c$. For a theoretical simulation of the QHE in constant current mode, we have to iterate Eq.~(\ref{eq:IHall}) with different values of $V_{\mathrm{SD}}$ until we match the desired total current.
In Fig.~\ref{fig:CC} we show the current dependence of the Hall curves for a magnetic field of ${\cal B}=10$~T as a function of Fermi energy, which can be adjusted experimentally by varying the voltage of a back-gate. The figures are calculated under the assumption, that a single electric field strength near the hot-spot dominates the current, which is 50 times stronger than the linear Hall field in the middle of the device. The increase of the local-electric field at the injection region of a factor 50 is an estimate based on the extension of observed hot-spots in GaAs/AlGaAs heterostructues \cite{Knott1995a}. For currents in the order of $10\;\mu$A a strong modulation of the Landau levels is visible which develop a very asymmetric shape and a gap near the filling factors $\nu=4,8,12,\ldots$, where the resistivity curves intersect at current independent points between two adjacent plateaus. These modulations and gaps are not caused by interactions but by the peculiar shape of the LDOS in graphene in the current injecting corner. We expect that these hot-spot induced gaps are reduced by the simultaneous emission from the other injection points with lower electric field values.\\
In conclusion, we have calculated the source-drain-voltage-current relation of a graphene device in strong magnetic fields. We have developed a theory of the QHE in graphene, which shows a rich substructure and modulation of Landau levels as a function of the back-gate voltage due to the presence of hot-spots in the device.
The modulation in graphene is different compared to other semiconductors due to the perfect alignment of the $n$th and $(n-1)$th Landau level around the same energy.
The experimental observation of current induced structures can clarify the electronic transport paths in graphene and pin down the region where electrons enter a device in the presence of high magnetic fields. The presence of disorder is not a requirement for the existence of the QHE.  Puddles of localized electrons away from the current injecting corner are not able to stop the injection process and thus may only deform the pathways of the electrons through the sample, which affect the longitudinal voltage drop, but not directly the Hall conductivity.\\
This work is supported by the Emmy-Noether program of the DFG (KR 2889-2/1). We thank P.~Kramer, C.~F.~Huang, L.-H.~Lin, J.~Fabian, and D.~Weiss for helpful discussions.
~\\

\end{document}